\documentclass[%
 aip,
 amsmath,amssymb,
 reprint,%
]{revtex4-1}

\usepackage{graphicx}
\usepackage{dcolumn}
\usepackage{bm}
\usepackage[mathlines]{lineno}
\usepackage{breqn}
\usepackage{amsmath}

\usepackage[utf8]{inputenc}
\usepackage[T1]{fontenc}
\usepackage{mathptmx}
\makeatletter
\let\cat@comma@active\@empty
\makeatother
\begin{document}
\preprint{AIP/123-QED}
\title{On the formalization of Asynchronous First Passage Algorithms}

\author{Luigi Sbail\`{o}}
\email{
luigi.sbailo@fu-berlin.de}
\author{Luigi Delle Site}
\email{luigi.dellesite@fu-berlin.de}
\affiliation{
Department of Mathematics and Computer Science,
Freie Universit\"{a}t Berlin,
Arnimallee 6,
D-14195,
Berlin,
Germany
}

\begin{abstract}
The formalization of First Passage schemes is revisited and the emerging
of a conceptual contradiction is underlined. We then show why, despite
such a contradiction, the numerical results are not explicitly affected.
Through a different formalization of the problem, we recast the current
principles of the algorithm in a more solid conceptual framework,
and numerical evidence gives further justification to our claims.
\end{abstract}
\maketitle
\section*{Introduction}

First passage algorithms have been widely used to efficiently simulate
the motion of random walkers in a lattice or Brownian particles in
a continuum. Applications, across different subjects and disciplines,
span from the determination of the ground state of bosonic particles
\cite{KLV} to the clustering of the oncoprotein Ras \cite{eGFRDras}
(further representative examples can be found in Refs.\cite{FPKMC,FPKMC2,FPKMC3,AccKMC,driftFPKMC,nucleationFPKMC,spatAnnFPKMC,eGFRD,eGFRDalldim,eGFRDtf,MD-GFRDanis}).
Although the algorithm has been successfully used for multiple numerical
applications, a proof of conceptual correctness with a solid formalism
has not been treated in full yet. 

The first passage algorithms allow
to efficiently simulate the free diffusion of particles assuming that
the motion of free particles is propagated with large spatio-temporal
jumps, sampled by appropriate first-passage Green's functions.
The volume of the system is decomposed into non overlapping domains, each
containing one or two particles and in every domain the underlying
stochastic dynamics is tractable analytically. The relevant consequence
is that the first exit-time of a single particle or of the center
of mass of a particles pair can be treated exactly. Particularly effective
at low molar concentrations, this approach can improve the computational
performance up to several orders of magnitude with respect to a time-driven
scheme \cite{MD-GFRD,MD-GFRDanis,MD-GFRDnew,FPKMC,FPKMC2,FPKMC3,eGFRDalldim}.
The algorithm was originally conceived in Ref.\cite{KLV} as a synchronous
scheme: (i) protective domains are constructed around each particle
and the corresponding exit-time is sampled; (ii) the shortest exit-time
is determined and particles are propagated to this time; (iii) finally,
new protective domains are constructed for all particles. Differently,
the first passage kinetic Monte Carlo algorithm (FPKMC) \cite{FPKMC}
is asynchronous (see Ref. \cite{asynch} for a more comprehensive discussion about asynchronous algorithms). 
In this scheme, the position of the particle involved
in the next exit-time event is updated locally, while the next events
of all other particles are maintained in a time ordered event-list,
which is updated every time a new event is sampled. The event-list
must be updated also for some unscheduled events, as, indeed, when
one particle is in proximity of another domain. In this case, in fact,
it is impossible to construct a domain of reasonable size without
overlapping with other domains. Then, the domain is burst, which means
that the particle position is updated inside the domain and the exit-time
previously sampled is removed from the event-list. More recently,
the Molecular Dynamics-Green's Function Reaction Dynamics (MD-GFRD)
algorithm \cite{MD-GFRD,MD-GFRDanis,MD-GFRDnew} has been developed.
This is an inherently multi-scale algorithm that pairs the event-driven
first exit-time sampling to a time-driven integration of the particles
motion. The particle interactions are locally integrated with an inter-particle
potential, while the free diffusion is updated with a first-passage
sampling. For simplicity, the above-mentioned schemes are called first
passage algorithms.

The event-based particle propagation is performed either
by placing the particle randomly on the domain borders when the pre-sampled
exit-time is reached, or by prematurely updating the particle position
inside the domain when this is burst before the exit-time. The probability
distribution that has been used when a domain is burst, is simply
given by the diffusion propagator with the imposed condition that
the particle has never reached the domain boundaries before the bursting-time
\cite{FPKMC,MD-GFRD}. The position sampling is performed after the
extraction of the first exit-time from the domain, but the information
of the exact moment when the particle hits the boundary is missing
as a condition for the probability distribution, although this would
be required \cite{Redner2}. Indeed, when an observable is sampled
at a time between the initial time and the time of a determined event,
the probability distribution should be conditional until realization
of the determined event. For instance, if the particle position is
sampled immediately before the exit-time, the particle is expected
to be in proximity of the domain borders. Let us assume that the sampling-time
$t$ is equal to the exit-time $\tau$ minus an infinitesimal time
$t=\tau-dt$, if the particle position $\vec{{r}}$ is sampled at
a finite distance $\Delta\vec{R}$ from the domain border $\vec{\Omega}$
then $\vec{r}=\vec{\Omega}-\Delta\vec{R}$, which leads to the evidence
that the particle travels a finite distance $\Delta\vec{R}$ in an
infinitesimal time $dt$. This is a contradiction, which might occur
when the condition on the exit-time is missing. Despite the conceptual
contradiction underlined above, numerical applications of first passage
schemes have always been shown to reproduce correctly the expected
results. The discussion above naturally leads to the need of a formal
proof of correctness of the algorithm which can clarify the issue
about the reported conceptual contradiction. In this paper, we investigate
the formal construction of the algorithm and clarify the pending aspects
of the previous discussion. Specifically, we show how the position
probability distribution used in previous first passage schemes can
be derived from the conditional probability distribution: the two
distributions result statistically identical when the bursting-time
is chosen randomly, as it happens in multi-particle simulations. 

The distinction between the conditional and unconditional
probability becomes relevant when a systematic bursting of the domains
is performed. This might be the case of MD-GFRD, a scheme where the
particles interaction is simulated with an inter-particle potential
in a time-discretization of the overdamped Langevin equation. When this approach
is used, the particles clock must be synchronized during their interaction
\cite{MD-GFRDnew}. This synchronization can be obtained by projecting
the exit-time onto a discrete temporal grid. The projection can be
performed after the exit-event with a fractional Brownian motion propagation
or before the exit-event by sampling the particle position in the
last discrete time before the exit-time. For the latter case, the
choice of the probability distribution is crucial, for the reason
that a systematic use of an unconditional probability distribution
under these circumstances would bring the simulation to inaccurate
results.

\section*{Results and Discussion}

\begin{figure}
\centering\includegraphics{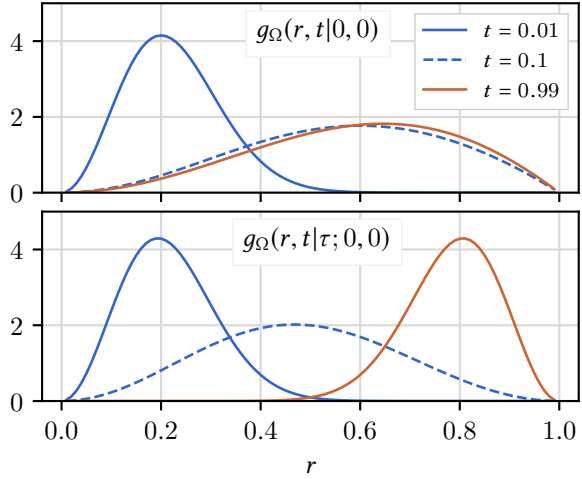}
\caption{Radial probability distributions as in Eq. (\ref{eq:pos_cond}) and
Eq. (\ref{eq:pos_uncond}). The domain $\Omega$ is a sphere centered
in the origin, and, given the spherical symmetry
of the problem, the probability distributions depend only on the radial
distance $r$ from the origin. All variables are adimensional. The
first exit-time has been given an unitary value $\tau =1 $; plots
have been performed for different values of the bursting-time $t$.
Immediately after the initial time the conditional and unconditional
probability distributions result identical. The particle is located
in proximity of the origin, and the information of the exit-time
is essentially irrelevant at this time. As time increases, the distributions diverge,
becoming clearly different when the bursting-time is close to the
extracted exit-time. }
\label{Fig:functPlot}
\end{figure}

The first exit-time from a protective domain $\Omega$ is derived
by imposing absorbing boundary conditions on the domain boundary $\partial\Omega$\cite{Redner}.
The Green's function $p_\Omega(\vec{r},t|\vec{r}_0,t_0)$ is the solution
to the diffusion equation when absorbing boundary conditions on $\partial\Omega$
and initial conditions in $\vec{r}_0$ at $t_0$ are imposed, and
it represents the probability that the particle is located in the
position $\vec{r}$ at time $t$ without having previously crossed the
domain borders. The integration of the probability distribution $p_\Omega(\vec{r},t|\vec{r}_0,t_{0})$
over the whole domain gives the survival probability $S_\Omega(t|\vec{r}_0,t_0)$,
i.e. the probability that the particle has not reached the domain
borders at $t$ yet. The first exit-time from the domain $q_\Omega(t|\vec{r}_0,t_0)$
is derived from the survival probability 

\begin{equation}
q_{\Omega}\left(t|\vec{r}_{0},t_{0}\right)=\frac{d\left(1-S_{\Omega}\left(t|\vec{r}_{0},t_{0}\right)\right)}{dt}.\label{eq:exit}
\end{equation}

It is assumed that the particle is initially located in the origin,
hence the exit-time is sampled from Eq. (\ref{eq:exit}) with $\vec{r}_0\rightarrow$0
at $t_0=0$. 

After the extraction of the particle escape at time $\tau$, the conditional
probability distribution of the particle position at an intermediate
time $0<t<\tau$ is obtained from the Bayes theorem

\begin{equation}
g_{\Omega}(\vec{r},t|\tau;0,0)=\frac{q_{\Omega}(\tau|\vec{r},t)p_{\Omega}(\vec{r},t|0,0)}{q_{\Omega}(\tau|0,0)}.\label{eq:pos_cond}
\end{equation}

In first passage schemes the unconditional probability is used, renormalized by the survival probability

\begin{equation}
g_{\Omega}(\vec{r},t|0,0)=\frac{p_{\Omega}(\vec{r},t|0,0)}{S_{\Omega}(t|0,0)}.\label{eq:pos_uncond}
\end{equation}

For a detailed derivation of the above probability distributions, we refer to the appendix \ref{App:derivations}.

\subsection*{Derivation of the unconditional probability}

Eq. (\ref{eq:pos_cond}) and Eq. (\ref{eq:pos_uncond}) are clearly
different (see Fig. \ref{Fig:functPlot}) and the assumption that
these two distributions can be used interchangeably might result surprising.
In particular in asynchronous schemes as FPKMC and MD-GFRD, the correctness
of the use of the unconditional probability is not evident, because
one may ask how the position probability of the particle inside the
domain is characterized when the position of other particles is updated
and known. This can be justified by considering the probability distribution
conditional with respect to the exit-time. Indeed, when the exit-time
is extracted, it is ensured that the particle is inside the domain
and that the particle position is probabilistically described by Eq.
(\ref{eq:pos_cond}) until realization of the exit-event.

It is possible to sample from Eq. (\ref{eq:pos_uncond}), only when
the sampling time is independent of the exit-time extraction. The
problem can also be reformulated by introducing $f(\vec{r},t|0,0)$,
which represents the unconditional probability that the particle is in $\overrightarrow{r}$
at $t$. The particle is supposed to be initially placed in a protective
domain $\Omega$. According to the Bayes theorem, the probability
distribution $f(\vec{r},t|0,0)$ can be decomposed in two conditional
probabilities with respect to the event that the particle never crossed
the domain borders $\partial\Omega$ until $t$

\begin{dmath}
f(\vec{r},t|0,0)=g(\vec{r},t|0,0;\vec{r}\in\Omega\thinspace\forall t^{\prime}\leq t)h(\vec{r}\in\Omega\thinspace\forall t^{\prime}\leq t)+
(1-h(\vec{r}\in\Omega\thinspace\forall t^{\prime}\leq t))\mathcal{P}(\vec{r},t|\partial\Omega),\label{eq:full}
\end{dmath}

where $h(\vec{r}\in\Omega\thinspace\forall t^{\prime}\leq t)$ is the probability
that the particle has never escaped the domain, and $\mathcal{P}(\vec{r},t|\partial\Omega)$ is the probability that the particle is in $\vec{r}$ after crossing the domain borders $\partial \Omega$. 
The particle position at $t$ can be sampled from the above equation as follows:

\begin{enumerate}
\item Sample the first exit-time $\tau$.
\item \textit{if} $t<\tau$:
\textit{then} the particle position is sampled from $g(\vec{r},t|0,0;\vec{r}\in\Omega\thinspace\forall t^{\prime}\leq t)$.
\item \textit{else if} $t>\tau$:
\textit{then} the particle position is the result of a Wiener process of time length $t-\tau$, starting in a random position on the domain borders.
\end{enumerate}

For a given observation time $t$ the first-exit time $\tau$ is extracted,
the comparison between these times, i.e. the $\it{if}$ statements in the list, in effect samples $h(\vec{r}\in\Omega\thinspace\forall t^{\prime}\leq t)$.
The only relevant assumption to sample $h(\vec{r}\in\Omega\thinspace\forall t^{\prime}\leq t)$ is that $t$ and $\tau$ are uncorrelated.

\begin{figure}
\centering\includegraphics{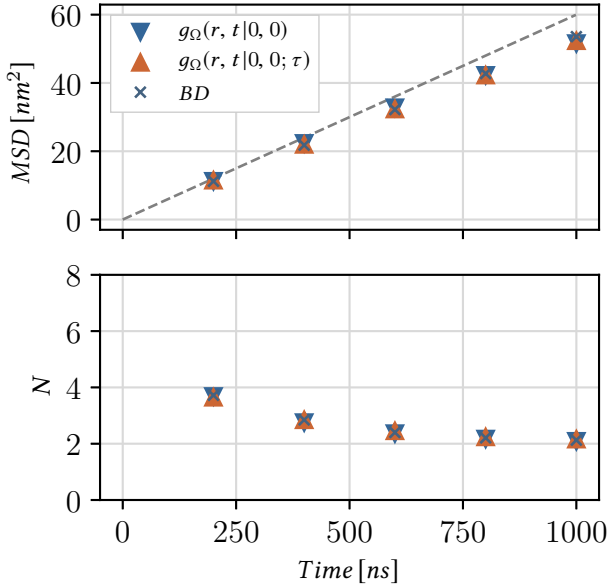}

\caption{Top: Mean squared displacement of Brownian particles simulated with the
hybrid MD-GFRD scheme as described in Ref. \cite{MD-GFRDnew} and with
a brute-force integration of the Brownian dynamics (BD). The hybrid
scheme was chosen to maximize the number of bursts
during simulations. The diffusion coeficient is $D=10\thinspace \mu m/s$
and the system volume is a box of length $14\thinspace nm$ with periodic
boundary conditions. Particles interact with a harmonic repulsive
potential when within a distance of $5\thinspace nm$. The dashed
line is the expected value for a free diffusion, BD lies below this slope for crowding effects.
Bottom: Kinetics of a two species annihilation reaction-diffusion system. The same simulation settings as above have been used. 
Particles are initially equally assigned to two different species $A,B$, and they annihilate when they collide with another particle of the same species,
 $A+A\rightarrow 0$ and $B+B\rightarrow 0$.
In both plots, the MD-GFRD simulations have been performed using conditional ($g_{\Omega}(r,t|\tau;0,0)$,
see Eq. (\ref{eq:pos_cond})) and unconditional ($g_{\Omega}(r,t|0,0)$,
see Eq. (\ref{eq:pos_uncond})) distribution probabilities to sample
the particle position at bursts, showing both a good agreement with
the BD integration. 
}
\label{Fig:P-PQ}
\end{figure}

\begin{figure}
\begin{center}
\includegraphics[scale=0.6]{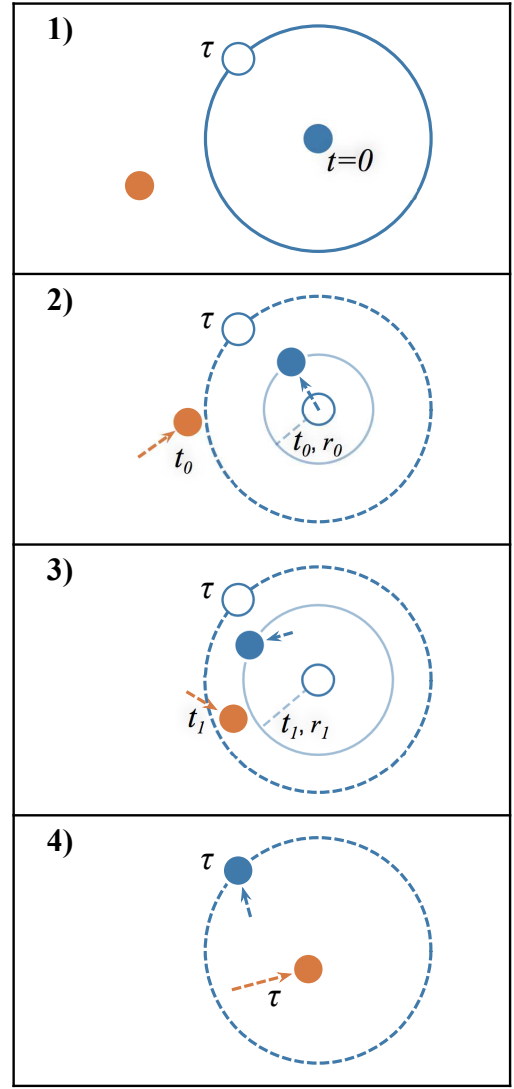}
\end{center}
\caption{Graphical representation of the motion of a particle after the bursting
of its domain. (1) The blue particle constructs a domain at time $t=0$,
and it samples $\tau$ as first exit-time from the domain. (2) The
orange particle bursts the domain at time $t_{0}$. The blue particle samples from Eq. (\ref{eq:pos_cond}) its position $r_{0}$
at $t_{0}$ and all successive free displacements that will take place
until the extracted exit-time $\tau$. The relative displacements
are recorded and successively used for simulating the free diffusion
of the particle until $\tau$. (3) The blue particle motion is integrated
in synchronous with the other particles. Given the linearity of the
Langevin equation, it is possible to integrate separately the free
diffusion component and the interaction with other particles. At each
discrete temporal step the free diffusion displacement is taken from
the relative displacements already sampled at the bursting-time, while
the displacements due to interactions are iteratively given by the
local configuration of the system. (4) In case of no previous interactions
the particle is exactly located on the domain borders at $\tau$.}

\label{Fig:scheme}
\end{figure}

The probability distribution $g(\vec{r},t|0,0;\vec{r}\in\Omega\thinspace\forall t^{\prime}\leq t)$
requires that the particle is known to be inside the domain at time
$t$ and that it has never left it before, conditions satisfied by
both Eq. (\ref{eq:pos_cond}) and Eq. (\ref{eq:pos_uncond}).
The connection between these probability distributions
is that Eq. (\ref{eq:pos_uncond}) is given by the integral of Eq.
(\ref{eq:pos_cond}) over all possible exit-times $\tau>t$ weighted
with their likelihood 

\begin{equation}
g_{\Omega}(\vec{r},t|0,0)=\int_{t}^{\infty}d\tau\thinspace g_{\Omega}(\vec{r},t|\tau;0,0)\thinspace\frac{q_{\Omega}(\tau|0,0)}{S_{\Omega}(t|0,0)},
\label{eq:marg}
\end{equation}

where $q_{\Omega}(\tau|0,0)/S_{\Omega}(t|0,0)$ is the probability
that the particle exits at $\tau$, conditional with respect to the
information that at $t$ is still inside the domain. From Eq. (\ref{eq:marg})
it is possible to infer that, given the information that the particle
at $t$ is still inside the domain, the position sampled from Eq.
(\ref{eq:pos_uncond}) is on average the same as if sampled from Eq.
(\ref{eq:pos_cond}), when the exit-time is extracted from Eq. (\ref{eq:exit}).
An evident assumption that has been made is that $t$ and $\tau$ are uncorrelated, 
otherwise the integral in Eq. (\ref{eq:marg}) could not be defined.
This assumption is the essential reason that explains why sampling from Eq. (\ref{eq:pos_uncond}) 
and from Eq. (\ref{eq:pos_cond}) are statistically equivalent, see  Eqs. (\ref{eq:full}) and (\ref{eq:marg}).

In Fig. \ref{Fig:P-PQ}, simulations have been performed using the two discussed probability distributions 
to update the particles position when domains are burst. 
Since in these simulations the domain bursts occur only in occasion of the random collision of an external particle with the protective domain of another particle,
it is reasonable to assume that the bursting-time and the first exit-time are uncorrelated, hence sampling from Eq. (\ref{eq:pos_uncond}) is statistically equivalent to 
sampling from  Eq. (\ref{eq:pos_cond}).
The mean squared displacement in a simple diffusive process and the kinetics of a two species annihilation reaction-diffusion system have been observed, and, as expected, the use of the conditional and 
unconditional distributions led to indistinguishable results.

\begin{figure}
\centering\includegraphics{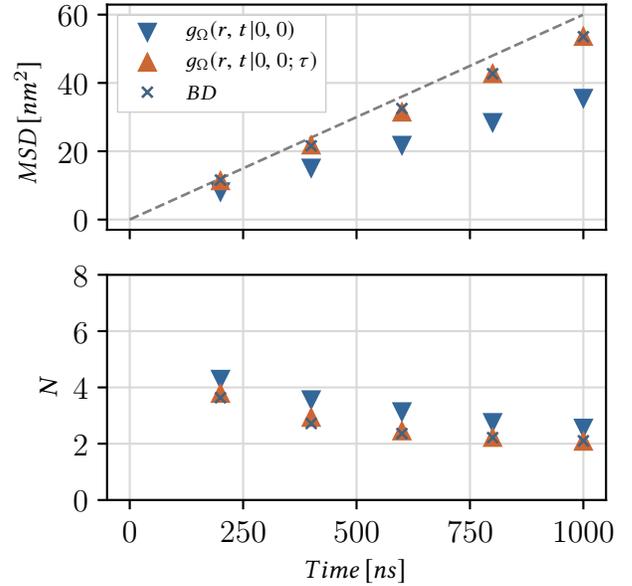}

\caption{Mean squared displacement of Brownian particles  and annihilation kinetics with the same properties
and in the same system as described in Fig. \ref{Fig:P-PQ}. In contrast
to Ref. \cite{MD-GFRDnew}, the clock synchronization is performed by sampling
the particle position at the last
discrete step before the exit-time. This procedure involves a systematic
domain bursting, where the bursting-time is correlated to the exit-time,
making the sampling from Eq. (\ref{eq:pos_uncond}) invalid. }

\label{Fig:proj}

\end{figure}

In asynchronous schemes as FPKMC and MD-GFRD
the particle position is updated serially while other particles lie
in protective domains, but it is still an open question how the motion
of particles in such protective domains is characterized before their
exit-time. Sampling the particles position from Eq. (\ref{eq:pos_uncond})
simply gives an average behavior that is justifiable when the bursting-time
is given randomly. More rigorously, the particle position should be
sampled from Eq. (\ref{eq:pos_cond}) until realization of the exit-time.
Let us suppose that a domain is burst at a time $t$ and the particle
that was inside the domain interacts with another particle on the
time grid $\ensuremath{\left\{ t_{i}\right\} _{i=0}^{n}}$, where
$t_{i+1}=t_{i}+dt$ and $dt$ is the integration step. The particle
position should be sampled from Eq. (\ref{eq:pos_cond}) at all time
steps until the pre-sampled exit time, $t>\tau$. However, in Eq.
$(\ref{eq:pos_cond}$) the domain has been assumed to be interaction-free
until $\tau$, but the particle is also assumed to be interacting.
It is still possible to use the conditional sampling because the Langevin
equation is linear, and the contribution of the free displacement
can be separated from the interaction term. When the particle is burst
at time $t$ all particle displacements until $\tau$ are sampled
and recorded. These will give the free displacements of the particle
that will then be summed to the interaction term (see Fig. \ref{Fig:scheme}). 
The sampling procedure just enlightened involves many integration steps between the bursting time and the
first-exit time, and in each of these steps the particle position is sampled from Eq. (\ref{eq:pos_cond}). 
This is clearly computationally more expensive than sampling once from Eq. (\ref{eq:pos_uncond}), 
that can be done efficiently using the rejection method as described in Refs.  \cite{FPKMC2,FPKMC3}

In the above derivations $t$ and $\tau$ have been assumed uncorrelated, therefore it is possible to sample from Eq.
$(\ref{eq:pos_uncond}$).
Although this assumption might result natural when the bursting-time
comes from the random encounters between different particles in a
multi-particle simulation, a systematic bursting happening at a time
which is function of the exit-time could be used in a multi-scale
simulation as MD-GFRD. If the particle synchronization, performed
in this multi-scale algorithm, is obtained by time-projection onto
the last discrete time before the exit-time of the particle, this
systematic bursting would clearly insert a correlation between $t$
and $\tau$. Eq. (\ref{eq:marg}) would not be valid because it is
not possible to integrate $\tau$ independently from $t$ and in Eq.
(\ref{eq:full}) $h(\vec{r}\in\Omega\thinspace\forall t^{\prime}\leq t)$ is not
sampled. 

In Fig. $\ref{Fig:proj}$, the mean squared displacements of particles
simulated as illustrated in Fig. \ref{Fig:scheme} and particles that use the position sampling
from Eq. ($\ref{eq:pos_uncond}$) are shown and compared to a brute-force
integration that is taken as reference. The synchronization step was performed before the exit-time during simulations, and it is clear from the figure that the use of Eq. ($\ref{eq:pos_uncond}$) makes the particles diffuse
slower. Indeed, immediately before the sampled exit-time the particle is expected to be in proximity of the domain borders, but this is not ensured without the condition on the exit-time (see Fig. $\ref{Fig:functPlot}$).  The correct slope of the mean square displacement is instead
reproduced when the Eq. ($\ref{eq:pos_cond}$) is sampled. In the
simulation, particles are interacting and their mean squared displacements
lie below the expected free diffusion value due to crowding effects. 

\section*{Conclusion}

We have analyzed the conceptual consistency of First Passage Algorithms. 
We assess that the use of the unconditional probability is only justified under the assumption that the bursting time and the first-exit time are uncorrelated. 
We also show that the unconditional probability distribution can be derived from a formally correct conditional probability distribution. 
The use of the unconditional probability distribution routinely employed leads to formal contradictions, although this aspect does not explicitly emerge in current numerical applications. 
Given the increasing number of algorithms which implement  asynchronous first passage schemes, we note that the enlightened contradiction might represent a pitfall in future applications.
We finally give numerical evidence that confirms the solidity of our formal derivation, and thus places First Passage Algorithms on firm conceptual ground. 

\appendix
\section{Probability distributions derivation}
\label{App:derivations}

The motion of a Brownian particle is assumed to be described stochastically by the diffusion equation

\begin{equation}
\frac{\partial}{\partial t}f(\vec{r},t)=D\Delta f(\vec{r},t),
\label{Eq:Diff}
\end{equation}

where $f(\vec{r},t)$ is the probability to find the particle in position $\vec{r}$ at time $t$, and $D$ 
is the diffusion coefficient. We assume that the particle position is known at time $t_0$, and we are interested in the first exit-time 
 of the particle from a domain $\Omega$ that is constructed around the particle. The domain $\Omega$ is supposed to be spherical 
 and centered on the particle position. Given the spherical symmetry of the system, the angular coordinates can 
be averaged out from Eq. (\ref{Eq:Diff})

\begin{equation}
f(\vec{r},t)=const\thinspace p(r,t).
\end{equation}

The first exit-time from a protective domain $\Omega$ is derived
by imposing absorbing boundary conditions on the domain boundary $\partial\Omega$\cite{Redner}.
The domain $\Omega$ is here assumed to be a sphere of radius $b$

\begin{equation}
p_{\Omega}(r=b,t)=0;\quad\forall t.
\label{Eq:abs_cond}
\end{equation}

This condition ensures that once the particle hits the domain borders, it is removed from the system. The integral 
of Eq. (\ref{Eq:abs_cond}) over the whole domain gives the survival probability $S_\Omega(t)$, i.e. the probability that, until $t$, the particle has always 
remained inside the domain 
 
 \begin{equation}
S_{\Omega}(t)=\int_0^b dr\thinspace 4\pi r^2\thinspace p_{\Omega}(r,t).
\label{Eq:S(t)}
 \end{equation}
 
The probability of crossing the domain borders for the first time at $t$ is derived from the survival probability 

\begin{equation}
q_{\Omega}\left(t|r_{0},t_{0}\right)=\frac{d\left(1-S_{\Omega}\left(t|r_{0},t_{0}\right)\right)}{dt}.\label{eq:exit_full}
\end{equation}

In this paper, our aim is to characterize the probability distribution of the particle position before it escapes the domain.
Firstly, the Green's function for the diffusion equation is found,  then exit-time boundary conditions 
are applied to the solution.
The diffusion equation is only solved for the radial component

\begin{equation}
\frac{\partial p(r,t)}{\partial t}=D\frac{1}{r^2}\frac{\partial}{\partial r} \left( r^2\frac{\partial p(r,t)}{\partial r}\right).
\end{equation}

The solution is found by assuming that it can be separated in the variables $t$ and $r$, $p(r,t)=g(r)h(t)$,

\begin{equation}
\frac{1}{D\thinspace h(t)}\frac{\partial h(t)}{\partial t} = \frac{1}{g(r)r^2}\frac{\partial}{\partial r} \left( r^2\frac{\partial g(r,t)}{\partial t}\right) .
\end{equation}

Each term of the above equation has only one variable, in this case both terms can be equated to a constant value resulting in two different equations

\begin{equation}
\frac{1}{D\thinspace h(t)}\frac{\partial h(t)}{\partial t} = \frac{1}{g(r)r^2}\frac{\partial}{\partial r} \left( r^2\frac{\partial g(r,t)}{\partial t}\right) =-\lambda^2.
\end{equation}

The solution of the first equation is an exponential function

\begin{equation}
h(t)=Ae^{-\lambda^2 D t}.
\label{Eq:h(t)}
\end{equation}

The second equation can be expressed as a harmonic oscillator differential equation by substituting $\tilde{g}(r)=r\thinspace g(r)$. 
The solution is then straightforward

\begin{equation}
g(r)=\frac{1}{r}\left(B\sin (\lambda r)+C\cos(\lambda r)\right).
\label{Eq:g(r)}
\end{equation}

The capital letters $A,B,C$ used are normalization factors.
Absorbing boundary conditions $p_\Omega(b,t)=0$ are imposed on the equation $p(r,t)=g(r)h(t)$ obtained from Eqs. (\ref{Eq:h(t)}) and (\ref{Eq:g(r)}) 

\begin{equation}
p_\Omega(r,t)=\frac{1}{r}\sum_m A_m e^{-\frac{m^2\pi^2}{b^2}Dt}\sin\left(\frac{m\pi}{b}r\right).
\end{equation}

The particle is known to be in $r_0$ at $t_0$

\begin{equation}
\begin{split}
p_\Omega&(r,t|r_0,t_0)=\frac{1}{2\pi b}\frac{1}{r\thinspace r_0}\cdot\\
&\cdot \sum_m e^{-\frac{m^2\pi^2}{b^2}D(t-t_0)}\sin\left(\frac{m\pi}{b}r\right)\sin\left(\frac{m\pi}{b}r_0\right).
\end{split}
\end{equation}

The survival probability is obtained as shown in Eq. (\ref{Eq:S(t)})

\begin{equation}
\begin{split}
S_\Omega&(t|r_0,t_0)=-\frac{2b}{\pi r_0}\cdot\\
&\cdot\sum_m \frac{(-1)^m}{m}  e^{-\frac{m^2\pi^2}{b^2}D(t-t_0)}\sin\left(\frac{m\pi}{b}r_0\right).
\end{split}
\end{equation}

The first exit-time is then derived from the survival probability (see Eq. (\ref{eq:exit_full}))

\begin{equation}
\begin{split}
q_\Omega&(\tau|r_0,t_0)=-\frac{2\pi D}{b r_0}\cdot\\
&\cdot\sum_m m(-1)^m e^{-\frac{m^2\pi^2}{b^2}D(t-t_0)}\sin\left(\frac{m\pi}{b}r_0\right).
\end{split}
\end{equation}

\bibliographystyle{aip}
\bibliography{all}

\end{document}